\documentclass[useAMS,usegraphicx,usenatbib,float]{mn2e}
\usepackage{color}
\usepackage{float}
\usepackage{graphicx}
\usepackage{natbib}
\usepackage{chngpage}
\usepackage{longtable}
\usepackage{amsmath}
\usepackage{amssymb}
\usepackage{mathrsfs}
\usepackage{multirow}
\usepackage[hang, raggedright, tight]{subfigure}

\def\simgt{\lower.5ex\hbox{\gtsima}}

\title{The catalogues and mid-infrared environment of Interstellar OH Masers }

\author[Haihua Qiao et al.]{{Haihua Qiao$^{1,2,3,4}$\thanks{E-mail:qiaohh@shao.ac.cn}, Juan Li$^{1,4}$, Zhiqiang Shen$^{1,3,4}$, Xi Chen$^{1,3,4}$, Xingwu Zheng$^{5,6}$}\\
$^{1}$Shanghai Astronomical Observatory, Chinese Academy of Sciences, 80 Nandan Road, Shanghai, China, 200030\\
$^{2}$University of Chinese Academy of Sciences, 19A Yuquanlu, Beijing, China, 100049\\
$^{3}$Key Laboratory for Research in Galaxies and Cosmology, Shanghai Astronomical Observatory, Chinese Academy of Sciences, 80 Nandan \\Road, Shanghai, China, 20030\\
$^{4}$Key Laboratory of Radio Astronomy, Chinese Academy of Science, 80 Nandan Road, Shanghai, China, 200030\\
$^{5}$School of Astronomy \& Space Science, Nanjing University, 22 Hankou RD, Jiangsu Province, China, 210093\\
$^{6}$Key Laboratory of Modern Astronomy and Astrophysics (Nanjing University), Ministry of Education, Jiangsu Province, China, 210093}

\begin{document}

%\date{Accepted . Received ; in original form }

%\pagerange{\pageref{firstpage}--\pageref{lastpage}} \pubyear{2010}

\maketitle

\label{firstpage}

\begin{abstract}

Data for a number of OH maser lines have been collected from surveys. The positions are compared to recent mid-infrared (MIR) surveys such as \emph{Spitzer}-GLIMPSE and \emph{WISE}, restricting the comparison to point sources. The colors and intensities of the IR sources are compared. There are many 18 cm OH masers, but far fewer in lines arising from higher energy levels. We also make a comparison with the 5 cm Class II methanol masers. We have divided the results into 3 subsamples: those associated with OH masers only, those associated with OH masers and Class II methanol masers, and those only associated with Class II methanol masers. There are no obvious differences in the color-color or color-magnitude results for the GLIMPSE point sources. However, according to the results from the \emph{WISE} 22 $\mu{}m$ survey, the sources associated with OH masers are brighter than those associated with methanol masers. We interpret the presence of OH and methanol masers mark the locations of regions where stars are forming. The OH masers are located on the borders of sharp features found in the IR. These are referred to as ¡°bubbles¡±. If the OH masers mark the positions of protostars, the result provides indirect evidence for triggered star formation caused by the expansion of the bubbles.

\end{abstract}

\begin{keywords}
methods: statistical - masers - stars: formation - ISM: bubbles.
\end{keywords}

\section{Introduction}

Interstellar Hydroxyl (OH) masers are an important tool for probing the environment of massive star-forming regions (SFRs). The maser phase is contemporaneous with the evolution of an ultra-compact (UC) H{\sc ii} region around the star (\cite{Re2002}), but dies out rapidly when the H{\sc ii}  region has enlarged to a size greater than 30 milliparsec (mpc) (\cite{Ca2001}). Thus the masers offer a means to discover the star at its early stage when it is hidden from the dust in the surrounding molecular clouds. Some recent interferometric observations, e.g. \cite{Mie2005}, \cite{Fie2007}, \cite{HC2007},  \cite{Bae2008}, and \cite{Sle2010}, provide high spectral resolutions and high positional accuracies, which are used to study the origins of the maser flares, maser velocity structures, and measure the magnetic strength in the SFRs. These studies shed new light on small-scale maser process and help us to understand the physical environment of the SFRs.

OH maser emission from the ground-state transitions ($^{2}\Pi_{3/2}$, $J = 3/2$ state) was firstly found toward several galactic H{\sc ii} regions (e.g. W3(OH)) by \cite{Wee1965} and \cite{Gu1965}. The 1665/1667 MHz ground-state transitions in SFRs are usually the strongest OH masers. They are generally accompanied by the weaker OH masers of other transitions, e.g. at excited transitions of $^{2}\Pi_{1/2}$, $J = 1/2$ and $^{2}\Pi_{3/2}$, $J = 5/2$. The excited state of OH ($^{2}\Pi_{1/2}$, $J = 1/2$ state) at 4765 MHz was firstly detected in the source W3(OH) and W49N (\cite{Zue1968}). The first detection of the excited state of OH ($^{2}\Pi_{3/2}$, $J = 5/2$ state) at 5 cm wavelength (6035 MHz) was made by \cite{Yee1969} toward W3. The highly excited state of OH ($^{2}\Pi_{3/2}$, $J = 7/2$ state) at 13441.417 MHz ($F = 4-4$) was discovered by \cite{Tue1970} in the W3(OH). The line radiation of excited OH (e.g. $^{2}\Pi_{1/2}$, $J = 1/2$ state; $^{2}\Pi_{3/2}$, $J = 5/2$ state; $^{2}\Pi_{3/2}$, $J = 7/2$ state) is particularly helpful in complementing the ground-state observations to understand the maser pumping process and interpret the physical conditions that are implied by the presence of the masers (\cite{Ca2001}). Extensive OH maser searches have been carried out toward color-selected infrared (IR) sources, or known SFRs associated with CH$_{3}$OH masers (e.g. Cohen et al. 1991; Cohen et al. 1995; Edris et al. 2007). These observations resulted in detections of hundreds of interstellar OH masers and supply an important tool for studies of maser pumping and physical conditions of their host SFRs (e.g. Szymczak et al. 2000; Szymczak \& G\'{e}rard 2004). However, until now there are no complete catalogues of all detected interstellar OH masers at each of the detected transitions. \cite{Mue2010} collected 3249 OH maser sources at 18 cm wavelength from the literature published up to April 2007, but the majority of the sources are stellar OH masers. Therefore, we performed an extensive literature search and compiled a complete catalogue of the detected interstellar OH masers so far.

The evolutionary phase that the interstellar OH masers trace the evolutionary sequence for different species of masers is still unclear now. \cite{El2006} examined the mid-infrared (MIR) properties of methanol masers with or without associated OH masers using the \emph{Spitzer} Galactic Legacy Infrared Mid-Plane Survey Extraordinaire (GLIMPSE) point source data. He proposed that OH masers may be generally associated with a later evolutionary phase or the stellar mass range associated with OH masers extends to higher masses than for methanol masers. However, because of the small numbers in each sample in \cite{El2006}, more data and investigations are needed to give firm conclusions. From this consideration, we make use of the GLIMPSE point source data to extensively study the MIR environment of the interstellar OH masers from larger size samples of OH and methanol masers. GLIMPSE is a legacy science program of the \emph{Spitzer} Space Telescope which covered the inner Galactic plane ($|l| \leq 65 ^{\circ}$) at 3.6, 4.5, 5.8, 8.0 $\mu{}m$ MIR wavelength bands with 1.4\arcsec-1.9\arcsec\ resolutions, with the Infrared Array Camera (IRAC; \cite{Bee2003}). As such, it offers us the best opportunity to study the MIR environment of interstellar OH masers and compare them with other sources. In addition, we also investigate the MIR environment of interstellar OH masers with the data from Wide-field Infrared Survey Explorer (\emph{WISE}) survey which mapped the full sky at 3.4, 4.6, 12 and 22 $\mu{}m$  with an angular resolution of 6.1\arcsec, 6.4\arcsec, 6.5\arcsec, 12\arcsec\ in the four bands, respectively (\cite{Wre2010}). \emph{WISE} can provide more information about the MIR environment of interstellar OH masers at longer MIR wavelengths which are complementary and important for our study. The \emph{Spitzer}-GLIMPSE images at 8.0 $\mu{}m$ revealed ``a bubbling Galactic disk" (\cite{Che2006}; \cite{Che2007}).  MIR bubbles are important and widespread morphological features in the interstellar medium (ISM). Bubbles could trigger the massive-star formation. During the expansion of the bubbles, neutral material accumulates on the border of bubbles, and becomes very massive with time. A new generation of stars may form in the collected layer (\cite{Dee2010}). In this case, OH masers are possibly associated with bubbles and might be found on the boarders of bubbles. Therefore, we investigate the relationship between OH masers and bubbles.

This paper is organised as follows. In Section 2, we introduce the data description. In Section 3, we describe the catalogues and present the relationship of flux densities between 18 cm OH masers and 5 cm OH masers. In Section 4, we discuss the IR environment of the interstellar OH masers, followed by a summary in Section 5.

\section{Data Description}

\subsection{Ground-state OH maser surveys}

Before the 1990's, ground-state OH maser surveys were mostly made with single-dish antennae, having rms position uncertainties greater than 10 \arcsec\ (e.g. \cite{Cae1980}; \cite{CH1983}; \cite{CH1987}). In the 1990's, interferometric OH maser surveys were carried out to obtain subarcsecond position accuracy. These observations include \cite{Ca1998} (hereafter C98), \cite{FC1999} (hereafter FC99), and \cite{Ca1999} (hereafter C99). Caswell (1998, 1999) used the Australia Telescope Compact Array (ATCA) to measure the positions for all the detected 18 cm OH masers before those two observaitons. With the Very Large Array (VLA), \cite{FC1999} presented the positions and spectra of 1665 MHz OH masers in 74 SFRs which were known to contain OH masers.
%table1%%%%%%%%%%%%%%%%%%%%%%%%%%%%%%%%%%%%%%

\begin{table*}
 \centering
\begin{minipage}{180mm}
\caption{\sffamily Summary of the interstellar OH maser surveys.
   \label{tab:tab1}}
\begin{tabular}{@{}llllllll@{}}
\hline
Code  &  Telescope  &  Reference  & $\#$det./$\#$obs.&  1$\sigma$[Jy]  & SN-RA(s)\footnote{ The significant decimal numbers for RA (s).} & SN-DEC(\arcsec)\footnote{ The significant decimal numbers for DEC (\arcsec).}& Selection  \\

\hline
18 cm & & & & & \\
\hline
C98 & ATCA & Caswell (1998) & -- & 0.04 & 2 & 1 & 18cm OH \\
FC99 & VLA & Forster et al. (1999) & -- & 0.1 & 2 & 1 & SFRs \\
C99 & ATCA & Caswell (1999) & -- & 0.03 & 2 & 1 & 18cm OH \\
A2000 & VLA & Argon et al. (2000) & 91/396 & 0.3 & 2 & 1 & 18cm OH,H$_{2}$O \\
SG2004 & Nan\c{c}ay & Szymczak et al. (2004) & 55/100 & 0.04 & 1 & 0 & Methanol \\
EF2007 & Nan\c{c}ay,GBT,100m & Edris et al. (2007) & 63/217 & 0.05--0.1 & 1 & 0 & IRAS(MYOs) \\

 \hline
 6 cm & & & & & \\
 \hline
CM91 & Lovell,76m & Gohen et al. (1991) & 9/44 & 0.05--0.3 & 1 & 0 & IRAS(18cm OH) \\
CM95 & Parkes,64m & Cohen et al. (1995) & 5/85 & 0.1 & 2 & 1 & IRAS,(U H{\sc ii} ) \\
S97 & HartRAO,26m & Smits (1997) & 6/29 & 0.3--0.5 & -- & -- & 6/18cm OH,Methanol \\
SK2000 & Torun,32m & Szymczak et al. (2000) & 9/57 & 0.7 & 2 & 1 & 18cm OH,H$_{2}$O \\
DE2002 & ATCA & Dodson et al. (2002) & 14/55 & 0.02 & 2 & 1 & SFRs \\
S2003 & HartRAO,26m & Smits (2003) & 3/69 & 0.2--0.4 & 1 & 0 & 18cm OH,Methanol \\
PG2004 & VLA & Palmer et al. (2004) & -- & $\sim$0.013 & 2 & 1 & 6cm OH \\
SC2005 & MERLIN & Harvey-Smith et al. (2005) & -- & 0.004--0.255 & 4 & 3 & SFRs \\

\hline
 5 cm & & & & & \\
 \hline
S94 & HartRAO,26m & Smits (1994) & 10/257 & $\sim$0.9 & -- & -- & OH,Methanol,H$_{2}$O  \\
CV95 & Parkes,64m & Caswell et al. (1995) & 52/208 & $\sim$0.1 & 1 & 0 & 18 OH \\
BD97 & Effelsberg,100m & Baudry et al. (1997) & 27/165 & -- & 2 & 1 & 18cm OH,H$_{2}$O ,IRAS \\
C2001 & ATCA & Caswell (2001) & -- & 0.04 & 2 & 1 & 5cm OH \\
C2003 & Parkes,64m & Caswell (2003) & -- & 0.03--0.05 & 2 & 1 & 5cm OH \\

\hline
 2.3 cm & & & & & \\
 \hline
BD2002 & Effelsberg,100m & Baudry et al. (2002) & 4/27 & $\sim$0.05 & 2 & 2 & 5cm OH \\
C2004 & Parkes,64m & Caswell (2004) & 8/56 & $\sim$0.022 & 2 & 1 & 5cm OH \\

\hline

\end{tabular}
\end{minipage}
\end{table*}
%table1%%%%%%%%%%%%%%%%%%%%%%%%%%%%%%%%%%%%%

In the 2000's, some targeted surveys were made by \cite{Are2000} (hereafter A2000), \cite{SG2004} (hereafter SG2004) and \cite{Ede2007} (hereafter EF2007) with single-dish or interferometric instruments. \cite{Are2000} observed 396 sources which have maser emission in OH and/or H$_{2}$O with NRAO 43 m telescope in Green Bank, obtained 91 interstellar OH masers with peak flux densities stronger than 1 Jy in both circular polarizations and finally mapped them with VLA. \cite{SG2004} made the observations of OH ground-state transitions in 100 methanol maser sources with the Nan\c{c}ay radio telescope, resulting in 55 OH maser detections. The detection rate of 55\% is the highest one in all targeted surveys for ground-state OH masers. \cite{Ede2007} observed 217 IRAS point sources exhibiting IR colors of high mass protostellar objects with the Nan\c{c}ay telescope and Green Bank Telescope (GBT), and detected 63 OH masers, 36 of which are new detections. Table \ref{tab:tab1} presents the detection rate, 1 - $\sigma$ sensitivity of the observations (when reported in the associated publication), the significant decimal numbers in RA (s) and DEC (\arcsec), and the target types for all ground-state and excited-state OH maser surveys.

\subsection{Excited-state OH maser surveys}

\subsubsection{6 cm OH maser surveys}

The targets for 6 cm OH maser surveys were generally selected from the known presence of 5/6/18 cm OH masers, H$_{2}$O masers, or methanol masers. \cite{Sze2000} (hereafter SK2000) carried out a survey towards 57 SFRs with 6/18 cm OH masers or methanol masers, and detected nine OH maser sources at 6 cm wavelength. \cite{Sm2003} (hereafter S2003) detected three OH masers at 6 cm wavelength from 69 SFRs with 18 cm OH masers or methanol masers. \cite{DE2002} (hereafter DE2002) carried out sensitive observations of OH maser emission at 6 cm wavelength towards 55 SFRs, and discovered fourteen sources.

Far-infrared flux has also been used to compile the survey samples. \cite{Coe1991} (hereafter CM91) chose 18 cm OH masers with strong IRAS 60 $\mu{}m$ flux density, together with eight known 6 cm OH masers as the survey sample, and detected nine OH masers at 6 cm wavelength. Subsequent work by \cite{Coe1995} (hereafter CM95) discovered five new OH masers at 6 cm wavelength from IRAS sources with FIR colors of UC H{\sc ii}  regions and embedded OB stars.

Some papers about monitoring the flux density, position measurements or phase-referenced observations for 6 cm OH masers were constructed. \cite{Sm1997} (hereafter S97) detected six OH masers at 6 cm wavelength from 29 SFRs with 6/18 cm OH masers or methanol masers, and monitored them for a period of more than a year. \cite{Pae2004} (hereafter PG2004) used VLA to determine the precise positions for 4765 MHz OH masers and the observations were successful for four OH masers. Recently, \cite{HC2005} (hereafter SC2005) made phase-referenced observations of thirteen SFRs at 4765 MHz with the Multi-Element Radio Linked Interferometer Network (MERLIN), and had ten detections at 4765 MHz, two detections at 4750 MHz, and one detection at 4660 MHz.

\subsubsection{5 cm OH maser surveys}

The targets for the 5 cm OH maser surveys were selected either by the presence of maser species (18 cm OH, H$_{2}$O, methanol masers) or by IRAS point sources exhibiting colors of UC H{\sc ii}  regions. \cite{Sm1994} (hereafter S94) searched for 6035 MHz OH masers towards 257 SFRs with 18 cm OH, H$_{2}$O, or methanol masers, and discovered ten new maser sources. \cite{CV1995} (hereafter CV95) detected 72 masers at the 6035 MHz transition towards 208 ground-state OH masers with 1665 MHz flux density greater than 0.8 Jy, and among them, 52 OH masers were firstly detected. \cite{Bae1997} (hereafter BD97) undertook a comprehensive search for 5 cm OH masers towards 165 sources, and detected 16 new OH masers.

Caswell (2001, 2003) (hereafter C2001, C2003) measured the positions of 5 cm OH masers with high accuracy and studied their polarization properties. He used ATCA to measure positions for more than 50 OH masers at the 6035 MHz transition with subarcsecond accuracy and discovered eleven new masers. He also studied the circular polarization properties in the spectra of 91 previously catalogued OH masers at the 6035 and 6030 MHz transitions with the Parkes radio telescope and provided more recent information for them.

\subsubsection{2.3 cm OH maser surveys}

 The 2.3 cm OH masers are usually searched towards 5 cm OH masers. \cite{BD2002} (hereafter BD2002) used the Effelsberg telescope to carry out sensitive observations towards 27 compact or ultra-compact H{\sc ii}  regions with 6035 MHz OH masers brighter than 0.5 Jy, and discovered three new masers at 13441 MHz in addition to W3(OH). \cite{Ca2004} (hereafter C2004) detected seven new masers at 13441 MHz from the sites of 56 catalogued 6035 MHz OH masers.

\section{Results}

\subsection{Catalogue description}

%table2
%%%%%%%%%%%%%%%%%%%%%%%%%%%%%%%%%%%%%%%%%%
\begin{table*}
\caption{\sffamily Catalogue of 18 cm interstellar OH masers. (This table (and Table 3, Table 4, Table 5) is available in its entirety in a machine-readable form in the online journal. A portion is shown here for guidance regarding its form and content.) \label{tab:tab2}}

\begin{tabular}{lllllcccll}
      %      \begin{minipage}{105mm}
\hline
\hline

No. &Source name & RA(J2000) & DEC(J2000)  & Freq.     & Vpeak         & Vrange      & Flux density&  Ref.  & Other name \\
 &            &   (h  m  s)        &($^{o}$  '  ")             & (MHz)     & (km s$^{-1}$) & (km s$^{-1}$) &   (Jy)&   &          \\

\hline

1 & G126.715-0.822 & 01 23 33.17 & +61 48 49.2 & 1665L & -12.06 & -12.06,-8.64 & $>$1.50 & A2000 & \\
1 & G126.715-0.822 & 01 23 33.17 & +61 48 49.2 & 1665R & -5.74 & & $>$0.50 & A2000 & \\
2 & G133.715+1.215 & 02 25 40.59 & +62 05 50.5 & 1665L & -39.29 & -43.96,-37.59 & $>$4.31 & A2000 & W3 \\
2 & G133.715+1.215 & 02 25 40.59 & +62 05 50.5 & 1665R & -39.2 & & $>$0.75 & A2000 & W3 \\
3 & G133.946+1.064 & 02 27 03.70 & +61 52 25.4 & 1612L & -43.15 & -47.51,-42.3 & $>$1.68 & A2000 & W3 OH \\
3 & G133.946+1.064 & 02 27 03.70 & +61 52 25.4 & 1612R & -42.17 & -42.92,-41.48 & $>$6.21 & A2000 & W3 OH \\
3 & G133.946+1.064 & 02 27 03.70 & +61 52 25.4 & 1665L & -46.29 & -48.94,-41.93 & $>$132.32 & A2000 & W3 OH \\
3 & G133.946+1.064 & 02 27 03.70 & +61 52 25.4 & 1665R & -45.02 & -48.9,-40.02 & $>$200.54 & A2000 & W3 OH \\
3 & G133.946+1.064 & 02 27 03.70 & +61 52 25.4 & 1667L & -44.43 & -47.77,-43.31 & $>$26.87 & A2000 & W3 OH \\

\hline
\end{tabular}

\end{table*}
%e\table2
%%%%%%%%%%%%%%%%%%%%%%%%%%%%%%%%%%

The detected interstellar OH masers and their basic information are listed in Tables 2, 3, 4, 5 (The full tables containing all sources are available electronically) for 18 cm, 6 cm, 5 cm, 2.3 cm OH masers, respectively. The catalogue entries are in order of the ascending right ascension (RA). The columns in Tables 2, 3, 4, 5 are as follows (the template is shown in Table 2): (1) the catalogue number; (2) the source name (collected from the literature); (3) and (4) the OH maser position in RA and declination (DEC) (J2000), respectively; (5) the observing frequency and its polarization information (some observations only have the single-polarization information); (6) the peak velocity of OH masers (km s$^{-1}$); (7) the velocity range of OH masers (km s$^{-1}$); (8) the flux density of OH masers (Jy; for the interferometer observations, the lower limit of the flux density is given); (9) the references; (10) the alias of OH masers.

The position accuracy may be inferred from the positions presented in the catalogues. The significant decimal numbers for RA (s) and DEC (\arcsec) are listed in Table \ref{tab:tab1}. The ATCA, VLA and some single-dish observations (C98, FC99, C99, A2000, CM95, SK2000, DE2002, PG2004, BD97, C2001, C2003, C2004) have two significant decimal numbers in RA and one in DEC. Positions from some single-dish measurements (SG2004, EF2007, CM91, S2003, CV95) have one significant decimal number in RA and none in DEC. \cite{BD2002} (BD2002) has two significant decimal numbers in both RA and DEC. The MERLIN observations (SC2005) have four and three significant decimal numbers in RA and DEC, respectively. Not all source positions have been determined with high accuracy, and this produces some difficulties when trying to draw firm conclusions about the associations.

The determination of the flux densities and velocities is described as follows. If the OH masers are observed with interferometers, such as A2000 and FC99 data, we choose the maximum flux density as the flux density of the source (the lower limit, marked with ``$>$") and its $V_{lsr}$ as the peak velocity. Then we determine the velocity range based on the maximum and minimum velocities of the source. If the source has both single-polarization observation and total flux density observation, the total flux density observation is selected. If the source has been observed for many times, the observation which has higher RA and DEC accuracy or gives the accurate flux density will be adopted. Sometimes, the source has several flux densities because of variability.

\subsection{The relationship between 1665 MHz flux density and 6035 MHz flux density}

\begin{figure}
  %\begin{center}
    \includegraphics[height=0.38\textwidth,width=0.48\textwidth, angle=0]{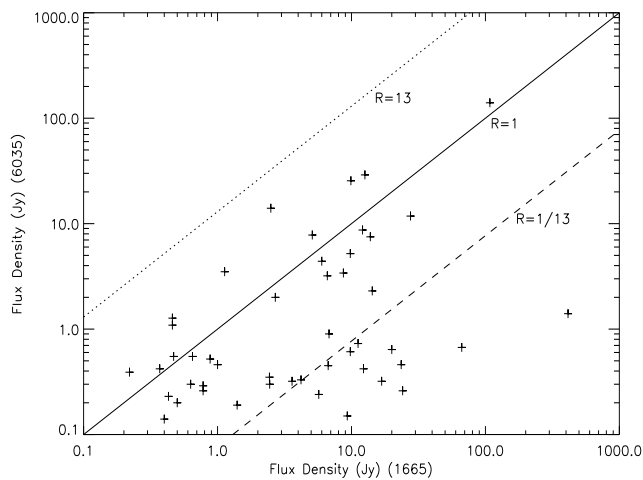}
  %\end{center}
  \caption{\sffamily
    Flux density of 1665 MHz transition vs. Flux density of 6035 MHz transition. 47 sources having both transitions are represented with crosses.
   \label{fig:fig1}}
\end{figure}

We use the 1665 MHz and 6035 MHz OH maser catalogues to investigate the relationship between the flux density of these two transitions. We select the OH masers, which have total flux densities (not the lower limit). In total, we have 249 sources at the 1665 MHz transition and 118 sources at the 6035 MHz transition. From \cite{FC1989}, \cite{Ca1997} and \cite{CaK2010}, if a separation between masers (assuming that they have the same distance) is greater than 6\arcsec, these masers are not closely associated (but probably in the same stellar cluster if their velocity ranges are similar).  So we take 6\arcsec\ as the separation criterion to investigate the association between these two samples, resulting in 47 sources which possess both transitions. Fig. \ref{fig:fig1} shows that there is a weak correlation between the flux density of the 1665 MHz transition and the 6035 MHz transition, which was expected by \cite{Gu1982}. The range of the flux density ratio $R=S(6035)/S(1665)$ is from 6 to $\frac{1}{400}$, and the majority of sources have the $R$ between 3 and $\frac{1}{6}$. \cite{CV1995} found that the greater the peak of 1665 MHz maser intensity is, the greater the detection rate of 6035 MHz masers is. \cite{PK1996} presented a detailed pumping scheme which accounts well for the ground-state OH masers (at 18 cm transition) in SFRs and can apparently also account quite well for the 5 cm transition under similar conditions (\cite{PK2000}). According to their model, the 6035 MHz line is weaker than the 1665 MHz line and is pumped towards higher gas densities compared to that for the 1665 MHz line.  The flux density ratio of 1665 MHz line and 6035 MHz line could be used to restrict the physical parameters of maser regions. For a line of frequency $\nu$ emitting from a source of area $\Omega$, the flux density $S$ is proportional to $T_{b}\nu^{2}\Omega$. In order to compare the observed flux density ratio $R=S(6035)/S(1665)$ with the model brightness temperature ratio ($T_{b}(6035)/ T_{b}(1665)$), an assumption about the relative size of the maser sources are required. If the 1665 MHz maser spots and 6035 MHz maser spots are comparable in size, the flux density ratios of $R=13$ and $R=1$ correspond to the brightness temperature ratio of unity and $\frac{1}{13}$. Based on \cite{Cre2002} model, we can roughly estimate the physical conditions of the maser region. For the gas kinetic temperature $T_{k}=30 K$, the dust temperature $T_{d}=175 K$, the flux density ratio $R=13$, the hydrogen density $n_{H}$ is $\sim10^{7.8} cm^{-3}$ and the specific column density $N_{OH}/\triangle V$ ($N_{OH}$ is the column density of OH; $\triangle V$ is the linewidth) is $\sim10^{11.85} cm^{-3}s$; For the same conditions of gas kinetic temperature and dust temperature, but with the flux density ratio $R=1$, the hydrogen density $n_{H}$ is $\sim10^{6.9} cm^{-3}$ and the specific column density $N_{OH}/\triangle V$ is $\sim10^{11.00} cm^{-3}s$.

% fig1
%%%%%%%%%%%%%%%%%%%%%%%%%%%%%%%
%\efi
%%%%%%%%%%%%%%%%%%%%%%%%%%%%%%%

\section{Discussion}

\subsection{GLIMPSE color}

\begin{figure}
  %\begin{center}
    \includegraphics[height=0.38\textwidth,width=0.48\textwidth, angle=0]{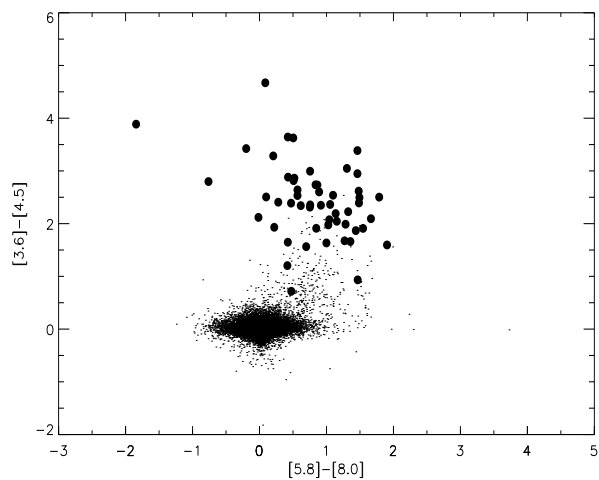}
  %\end{center}
  \caption{\sffamily
Color-color diagram constructed from GLIMPSE point source catalogue. The OH masers with GLIMPSE counterparts are represented with solid circles. Sources within 30\arcmin\ radius of $l=320.0^{\circ}, b=0.0^{\circ}$ are represented with dots. Only sources for which there is flux density information for all four IRAC bands have been included in the plot, that is, 54 of 113 OH masers with a GLIMPSE point source within 2\arcsec\, and 20018 of 101615 of the comparison sample.
   \label{fig:fig2}}
\end{figure}

We have plotted color-color and color-magnitude diagrams for the GLIMPSE point sources associated with the interstellar OH masers which have better position accuracies with ATCA or VLA observations. For comparison, we choose the GLIMPSE sources within 30\arcmin\ radius of $l=320.0^{\circ}$, $b=0.0^{\circ}$. The comparison sample from the GLIMPSE point source catalogue includes 101,615 sources and provides enough information to investigate the color and magnitude difference between maser-associated GLIMPSE sources and normal GLIMPSE sources. The interstellar OH masers are from C98 with ATCA observations and A2000 with VLA observations. The former sample includes 206 interstellar OH masers and the latter sample contains 91 interstellar OH masers. Then we associate the positions of these two samples taking 2\arcsec\ as the separation criterion and obtain 266 interstellar OH masers. The positional accuracy of these 266 interstellar OH masers is about 0.4\arcsec. Among these 266 interstellar OH masers, 219 OH masers are in the GLIMPSE survey region ($|l| \leq 65 ^{\circ}$), and 113 OH masers are associated with GLIMPSE point sources within a 2\arcsec radius. The maser-associated GLIMPSE sources are obviously offset from the majority of the comparison sources in the color-color and color-magnitude diagrams and have much redder colors. The conclusion can be easily drawn from Fig. \ref{fig:fig2}, which is a plot of the [5.8]-[8.0] versus [3.6]-[4.5] colors, showing that the maser-associated sources mostly lie above the vast majority of the comparison sources. This result is similar to the color-color diagram of methanol masers (see Figure 15 of Ellingsen 2006).

% fig2
%%%%%%%%%%%%%%%%%%%%%%%%%%%%%%%
%\efi
%%%%%%%%%%%%%%%%%%%%%%%%%%%%%%%

The sources associated with interstellar OH masers lie in a distinctive region of GLIMPSE color-color diagrams and occupy the same area as methanol masers. This result is interesting, but what can we derive from it? As discussed by \cite{El2006}, polycyclic aromatic hydrocarbon (PAH) emission and silicate absorption lines can complicate the problem and affect some of IRAC bands (see \cite{Dr2003} for a detailed discussion). But the actual influence of PAH emission and silicate absorption on the observed IRAC colors needs to be confirmed and measured by MIR spectroscopy. For this reason the interpretation of GLIMPSE colors of maser-associated sources is uncertain to a large degree. Nevertheless, it is still meaningful to make the comparison between the GLIMPSE point sources with and without OH masers, because they all suffer the same effects and limitations with the same instrument.

% fig3
%%%%%%%%%%%%%%%%%%%%%%%%%%%%%%%
\begin{figure}
  %\begin{center}
    \includegraphics[height=0.38\textwidth,width=0.48\textwidth, angle=0]{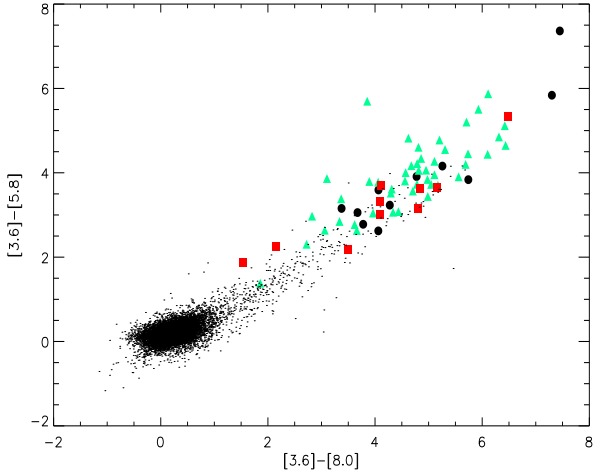}
  %\end{center}
  \caption{\sffamily
GLIMPSE [3.6]-[8.0] vs. [3.6]-[5.8] color-color diagram. The OH masers with associated GLIMPSE point sources are represented with solid circles if they don't have associated methanol multibeam (MMB) masers, as solid triangles if they have. The solid squares represent the methanol masers with associated GLIMPSE point sources but without associated OH masers, and these methanol masers are from Ellingsen (2006). Sources within 30\arcmin\ radius of $l=320.0^{\circ}, b=0.0^{\circ}$ are represented with dots.
   \label{fig:fig3}}
\end{figure}
%\efi
%%%%%%%%%%%%%%%%%%%%%%%%%%%%%%%

% fig4
%%%%%%%%%%%%%%%%%%%%%%%%%%%%%%%
\begin{figure}
  %\begin{center}
    \includegraphics[height=0.38\textwidth,width=0.48\textwidth, angle=0]{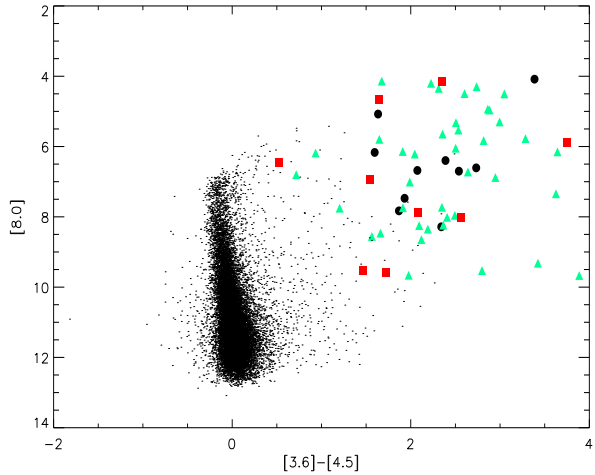}
  %\end{center}
  \caption{\sffamily
GLIMPSE [3.6]-[4.5] vs. 8.0 $\mu{}m$ color-magnitude diagram. The OH masers with associated GLIMPSE point sources are represented with solid circles if they don't have associated methanol multibeam (MMB) masers, as solid triangles if they have. The solid squares represent the methanol masers with associated GLIMPSE point sources but without associated OH masers, and these methanol masers are from Ellingsen (2006). Sources within 30\arcmin\ radius of $l=320.0^{\circ}, b=0.0^{\circ}$ are represented with dots.
   \label{fig:fig4}}
\end{figure}
%\efi
%%%%%%%%%%%%%%%%%%%%%%%%%%%%%%%

 The modeling of \cite{Whe2003} shows that Class 0 objects lie in the same area in the GLIMPSE [5.8]-[8.0] versus [3.6]-[4.5] color-color diagram as the interstellar OH masers and the methanol masers. It is interesting to see whether or not OH masers trace the same evolutionary stage as methanol masers. The methanol multibeam (MMB) survey is a blind survey with longitudes from $186^{\circ}$ to $20^{\circ}$ and includes 707 6.7 GHz Class II methanol masers in total (\cite{Cae2010}; \cite{Gre2010}; \cite{Cae2011}; \cite{Gre2012}). We cross-match the positions of 707 MMB masers and 113 interstellar OH masers which are associated with GLIMPSE point sources within 2\arcsec. Finally we obtain 79 OH masers which are also associated with methanol masers and 25 OH masers which are not associated with methanol masers (solitary OH masers for short). Besides we collect 35 methanol masers without an associated OH maser (solitary methanol masers for short) from Table 1 of Ellingsen's (2006) paper. The sample of his paper is a statistically complete sample detected in the Mt.Pleasant survey (\cite{Ele1996}) and includes 56 methanol masers. Among the 35 solitary methanol masers, only 17 methanol masers have GLIMPSE counterparts. We construct color-color and color-magnitude diagrams for these three samples. Fig. \ref{fig:fig3} plots a [3.6]-[8.0] versus [3.6]-[5.8] color-color diagram, using different symbols for the three samples of maser-associated GLIMPSE sources. The comparison sources are also shown with dots. Fig. \ref{fig:fig3} shows that the colors of the solitary methanol masers tend to be little bluer than that of the solitary OH masers, but not obviously, and the OH masers with an associated methanol maser lie between them. Fig. \ref{fig:fig4} shows the [3.6]-[4.5] color and 8.0 $\mu{}m$  magnitude diagram. There is no obvious difference among the solitary methanol masers, the OH masers with an associated methanol maser and the solitary OH masers. This result suggests that these two masers are possibly tracing closely evolutionary stage and cannot be clearly distinguished from the GLIMPSE color-color and color-magnitude diagrams, which is different from Figure 19 of \cite{El2006} which shows that methanol masers with associated OH masers are generally brighter at 8.0 $\mu{}m$ than those without. Perhaps the data is not large enough to reveal the color and magnitude difference between the two species of masers. Thus it is still necessary and interesting to compare the MIR properties of the GLIMPSE sources associated with OH masers and methanol masers in a larger and more complete sample in order to investigate maser evolutionary sequence.

\subsection{\emph{WISE} colors}

\begin{figure}
  %\begin{center}
    \includegraphics[height=0.38\textwidth,width=0.48\textwidth, angle=0]{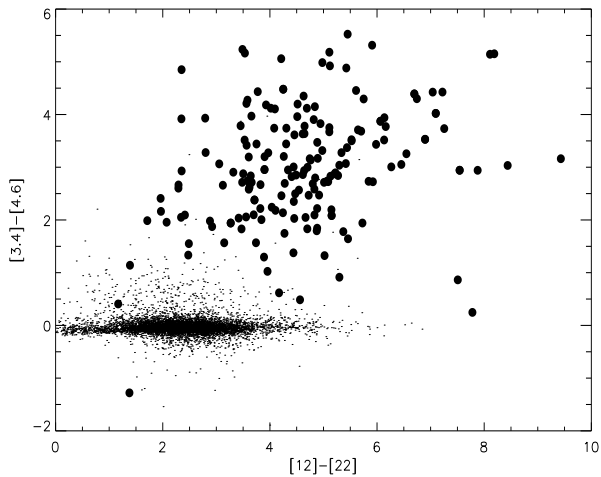}
  %\end{center}
  \caption{\sffamily
Color-color diagram constructed from \emph{WISE} point source catalogue. The OH masers with \emph{WISE} counterparts are represented with solid circles. Sources within 20\arcmin\ radius of $l=300^{\circ}, b=0^{\circ}$ are represented with dots. Only sources for which there is flux density information for all four bands have been included in the plot, that is, 181 of 205 OH masers with a \emph{WISE} point source within 6\arcsec\, and 7728 of 7729 of the comparison sample.
   \label{fig:fig5}}
\end{figure}

 Since \emph{WISE} has lower angular resolution than GLIMPSE, we take 6\arcsec\ as the criterion to search for the \emph{WISE} counterparts for 266 interstellar OH masers. Among 266 OH masers, 205 OH masers have \emph{WISE} counterparts. After cross-matching them with 707 MMB masers, we obtain 32 OH masers without an associated methanol maser (solitary OH masers), which also lie in the MMB survey region. We get 139 OH masers which have associated methanol masers. We also search for the \emph{WISE} counterparts for the 35 methanol masers without an associated OH maser (solitary methanol masers) from Table 1 of Ellingsen's paper (2006), and we find 23 \emph{WISE} counterparts within 6\arcsec. For comparison, we select the \emph{WISE} sources within 20\arcmin\ radius of $l=300^{\circ}$, $b=0^{\circ}$. The comparison sample includes 7729 sources and has enough color and magnitude information for detailed MIR environment study. Then we construct the color-color and color-magnitude diagrams for these four samples including comparison sample to investigate the MIR environment of masers. The maser-associated \emph{WISE} sources are clearly offset from the vast majority of the comparison sources in most of the color-color and color-magnitude diagrams, and their colors are much redder. These features are the same as their GLIMPSE features. This can be clearly seen in Fig. \ref{fig:fig5}, which plots the [12] - [22] versus [3.4] - [4.6] color-color diagram, showing the deviation between the maser-associated \emph{WISE} sources and the comparison sources.

% fig5
%%%%%%%%%%%%%%%%%%%%%%%%%%%%%%%
%\efi
%%%%%%%%%%%%%%%%%%%%%%%%%%%%%%%

% fig6
%%%%%%%%%%%%%%%%%%%%%%%%%%%%%%%
\begin{figure}
  %\begin{center}
    \includegraphics[height=0.38\textwidth,width=0.48\textwidth, angle=0]{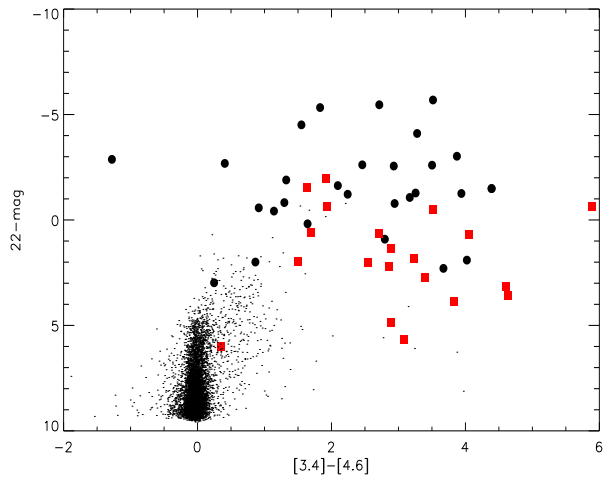}
  %\end{center}
  \caption{\sffamily
\emph{WISE} [3.4]-[4.6] vs. 22 $\mu{}m$ color-magnitude diagram. The OH masers with associated \emph{WISE} point sources are represented with solid circles if they do not have associated methanol multibeam (MMB) masers. And the methanol masers with associated \emph{WISE} point sources are represented with solid squares if they don't have associated OH masers. \emph{WISE} point sources within 20\arcmin\ radius of $l=300^{\circ}, b=0^{\circ}$ are represented with dots.
   \label{fig:fig6}}
\end{figure}
%\efi
%%%%%%%%%%%%%%%%%%%%%%%%%%%%%%%

% fig7
%%%%%%%%%%%%%%%%%%%%%%%%%%%%%%%
\begin{figure}
  %\begin{center}
    \includegraphics[height=0.38\textwidth,width=0.48\textwidth, angle=0]{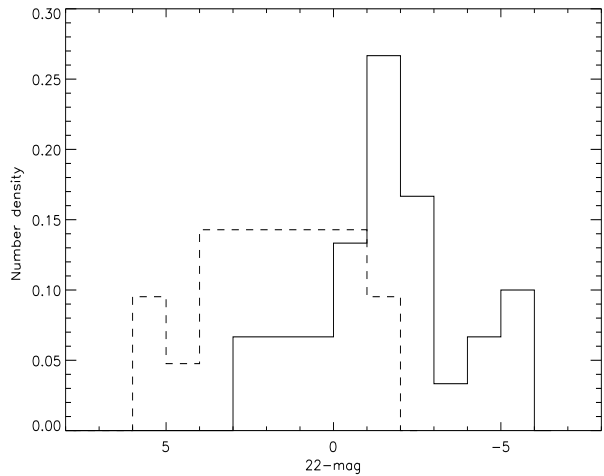}
  %\end{center}
  \caption{\sffamily
The distribution of 22 $\mu{}m$ magnitude. The dashed curve shows the distribution of the methanol masers without associated OH masers, and the solid curve is that of the OH masers without associated methanol masers.
   \label{fig:fig7}}
\end{figure}
%\efi
%%%%%%%%%%%%%%%%%%%%%%%%%%%%%%%

% fig8
%%%%%%%%%%%%%%%%%%%%%%%%%%%%%%%
\begin{figure}
  %\begin{center}
    \includegraphics[height=0.38\textwidth,width=0.48\textwidth, angle=0]{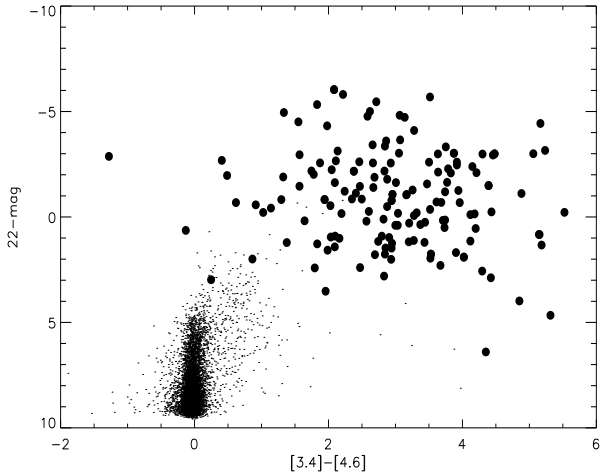}
  %\end{center}
  \caption{\sffamily
\emph{WISE} [3.4]-[4.6] vs. 22 $\mu{}m$ color-magnitude diagram for the OH masers same as Fig. \ref{fig:fig5}. The OH masers with \emph{WISE} counterparts are represented with solid circles. Sources within 20\arcmin\ radius of $l=300^{\circ}, b=0^{\circ}$ are represented with dots.
   \label{fig:fig8}}
\end{figure}
%\efi
%%%%%%%%%%%%%%%%%%%%%%%%%%%%%%%

% fig9
%%%%%%%%%%%%%%%%%%%%%%%%%%%%%%%
\begin{figure}
  %\begin{center}
    \includegraphics[height=0.38\textwidth,width=0.48\textwidth, angle=0]{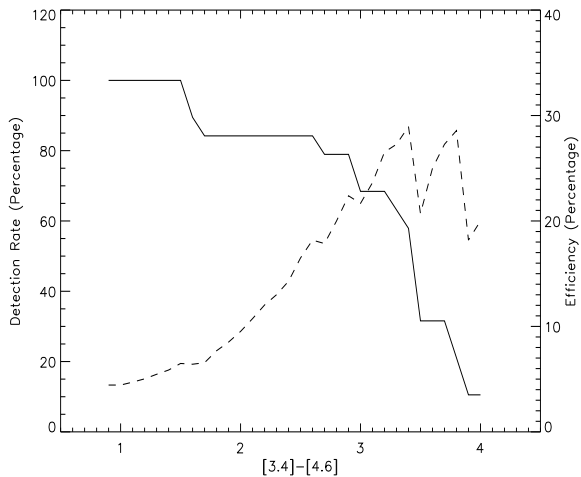}
  %\end{center}
  \caption{\sffamily
\emph{WISE} [3.4]-[4.6] color vs. Detection Rate/Efficiency diagram (22 $\mu{}m$ magnitude $<3$). The Detection Rate is represented with the solid line. And the Efficiency is represented with the dashed line.
   \label{fig:fig9}}
\end{figure}
%\efi
%%%%%%%%%%%%%%%%%%%%%%%%%%%%%%%

Fig. \ref{fig:fig6} plots the color-magnitude diagram of [3.4] - [4.6] color versus 22 $\mu{}m$ magnitude, which shows that the solitary OH masers have a brighter MIR source counterpart at 22 $\mu{}m$, compared to the solitary methanol masers. We also plot the histograms of the four bands for the \emph{WISE} sources associated with solitary OH masers and solitary methanol masers. We find that at 4.6, 12, 22 $\mu{}m$, there is a trend that the \emph{WISE} sources associated with solitary OH masers are brighter than the \emph{WISE} sources associated with solitary methanol masers, and it is most obvious at 22 $\mu{}m$ band (shown in Fig. \ref{fig:fig7}). This result may have a bias because the 22 $\mu{}m$ magnitude depends on the distance. However, the \emph{WISE} sources associated with these two kinds of masers all suffer from the same effect, and the obvious tendency could still illustrate some problems. The 22 $\mu{}m$ band should trace the same dust emission components as the 24 $\mu{}m$ band, which likely comes from very small grains (VSGs) out of thermal equilibrium, or a combination of emission from VSGs and from a larger grain population whose temperature is closer to 25 K (\cite{Ane2012}). The OH masers with the brighter 22 $\mu{}m$ emission may trace a later evolutionary stage of the central star than the methanol masers because of the higher dust temperature. \cite{Cre2002} found that the gas-phase molecular abundance is the key determinant of maser activity for both CH$_{3}$OH and OH masers. A large CH$_{3}$OH column density can be easily reached in SFRs due to a high abundance of methanol ice on grain mantles (\cite{Dae1999}). OH abundance can be enhanced by the photodissociation or ion-molecule process after the H$_{2}$O molecules are injected into the gas phase. Charnley et al. (1992, 1995) predicted that a maximum of methanol abundance appears before a peak of OH abundance. If these models are confirmed by further chemical modeling, this would support the proposed evolutionary sequence mentioned above. Alternatively, the stellar mass range associated with OH masers may extend to higher masses than that for methanol masers as revealed by \cite{El2006}.

The distinctive color-color and color-magnitude properties of the \emph{WISE} sources associated with OH masers provide an opportunity to create a \emph{WISE}-selected target sample for future OH maser searches. Fig. \ref{fig:fig8} is the [3.4]-[4.6] vs. 22 $\mu{}m$ color-magnitude diagram, showing that the majority ($\sim$80\%) of the \emph{WISE} sources associated with known OH masers locate in a domain of $[3.4]-[4.6]>2$ and 22 $\mu{}m$ magnitude $<3$, while nearly null sources in the comparison field locate in this domain. On the other hand, it is possible to estimate the detection rate and efficiency of an OH maser search targeted toward \emph{WISE} point sources by comparing with one blind or untargeted OH maser survey. One untargeted survey with the Parkes by Caswell et al. (1980) had detected 19 interstellar OH masers in the region of $l=330^{\circ}-340^{\circ}$, $b=-0.3^{\circ}$ to $0.3^{\circ}$ at a sensitivity of $\sim$0.1 Jy. For a search with a single dish, any targets that lie within half the FWHM beam could be searched only once in a single pointing. Considering the spatial coverage of the Parkes beam ( $\sim$10 \arcmin\ at 1.66 GHz), the actual targeted \emph{WISE} sources could be deduced. Based on these, we plot the dependence of the detection rate and efficiency with the [3.4]-[4.6] color (under 22 $\mu{}m$ magnitude $<3$) in Figure 9, using the \emph{WISE} sources and the OH maser data in the Caswell et al. (1980) surveyed region. From this figure, for the \emph{WISE} sources satisfying the criteria of $[3.4]-[4.6]>2$ and 22 $\mu{}m$ magnitude $<3$, 84\% (16/19) of known OH masers in the blind survey would be expected to detect, this detection rate/percent is consistent with that (80\%) shown in Fig. \ref{fig:fig8} for all known OH masers. Therefore, this also suggests that it is more reasonable to estimate a detection efficiency of OH maser searches using the criteria of $[3.4]-[4.6]>2$ and 22 $\mu{}m$ magnitude $<3$, under which a detection efficiency of $\sim$10\% would be achieved. We searched for the \emph{WISE} All-sky Data Release and found about 7559 point sources satisfying the criteria. These \emph{WISE} sources would provide a potential target sample for the further OH maser searches, especially with the newly-built 65 meter radio telescope in Shanghai. It should be noted that this new telescope has a similar beamsize to the Parkes dish. As a simple estimation, considering the beam coverage of the Shanghai 65 meter or Parkes telescope, the target pointing positions are reduced to 5209 towards these \emph{WISE}-selected sources, thus $\sim$500 ground-state interstellar OH masers would be expected to detect. The expected total number of ground-state interstellar OH masers would be $\sim$600 in our Galaxy after considering that 84\% of all OH masers would be detected from the above statistics.

\subsection{Bubble-Like structures}

 \begin{table*}

 \leftline{\textbf{Table 6. }Maser-associated bubbles.}
  \label{tab:tab6}
    \vspace{8pt}
   % \begin{center}
  %    \begin{minipage}{105mm}
      \begin{tabular}{lrrcccccll}

    \hline
    \hline
\scriptsize
Catalogue No.	&	$l$	&	$b$		&	$R_{out}$	&	Eccentricity	&	$\langle{R}\rangle$	&	$\langle{T}\rangle$	&	Morphology Flags	& Name (OH) &Name (CH3OH)\\

	&	(deg)	&	(deg)		&	(arcmin)	&		&	(arcmin)	&	(arcmin)	&	&&	\\

            \hline

N2 & 10.747  & -0.468 & 8.31  & 0.56  & 6.95  & 1.23  & B & G10.623-0.383 &  \\
N65 & 35.000  & 0.332 & 2.59  & 0.49  & 2.15  & 0.54  & C & G35.024+0.350 &  \\
N68 & 35.654  & -0.062 & 6.06  & 0.72  & 4.68  & 0.74  & C,CC & G35.577-0.029 &  \\
S36 & 337.971  & -0.474 & 3.56  & 0.66  & 2.73  & 0.69  & C & G337.916-0.477 &  \\
S62 & 331.316  & -0.359 & 2.50  & 0.58  & 2.02  & 0.45  & B & G331.342-0.346 & G331.342-0.346 \\
S66 & 330.781  & -0.414 & 7.42  & 0.69  & 5.63  & 1.37  & B,CC,MB & G330.878-0.367 &  \\
 &  &  &  &  &  &  &  & G330.878-0.367 &  \\
S110 & 316.809  & -0.031 & 1.82  & 0.63  & 1.42  & 0.36  & C,TP & G316.811-0.057 & G316.811-0.057 \\
S122 & 313.418  & 0.128 & 8.69  & 0.78  & 6.22  & 1.31  & C & G313.469+0.190 & G313.469+0.190 \\
S169 & 301.122  & -0.152 & 4.06  & 0.34  & 3.40  & 1.07  & B & G301.136-0.226 & G301.136-0.226 \\
CN15 & 0.562  & -0.843 & 1.03  & 0.73  & 0.71  & 0.27  & B, BC & G0.546-0.852 &  \\
CN71 & 5.894  & -0.463 & 6.17  & 0.38  & 5.45  & 0.96  & B, FL, Y & G5.886-0.393 &  \\
 &  &  &  &  &  &  &  & G5.885-0.392 &  \\
CN135 & 9.612  & 0.196 & 0.48  & 0.69  & 0.34  & 0.13  & C, BC & G9.619+0.193 & G09.619+0.193 \\
 &  &  &  &  &  &  &  & G9.620+0.194 &  \\
 &  &  &  &  &  &  &  & G9.621+0.196 & G09.621+0.196 \\
CS1 & 359.965  & -0.502 & 2.74  & 0.67  & 2.15  & 0.42  & B, Y & G359.970-0.457 & G359.970-0.457 \\
CS43 & 354.715  & 0.297 & 0.51  & 0.70  & 0.38  & 0.11  & C, Y & G354.724+0.300 & G354.724+0.300 \\
CS46 & 354.619  & 0.492 & 1.25  & 0.50  & 1.04  & 0.24  & C, BC, Y & G354.615+0.472 & G354.615+0.472 \\
CS55 & 353.417  & -0.375 & 1.34  & 0.91  & 0.78  & 0.15  & B & G353.410-0.360 & G353.410-0.360 \\
CS73 & 352.174  & 0.297 & 6.80  & 0.62  & 5.52  & 1.02  & B & G352.161+0.200 &  \\
CS106 & 350.327  & 0.096 & 1.02  & 0.75  & 0.74  & 0.18  & B, CS & G350.329+0.100 &  \\

             \hline

      \end{tabular}
 %  \end{minipage}
  %\end{center}

\end{table*}

% fig10
%%%%%%%%%%%%%%%%%%%%%%%%%%%%%%%
\begin{figure}
  %\begin{center}
    \includegraphics*[height=0.5\textwidth,viewport=320 20 900 650]{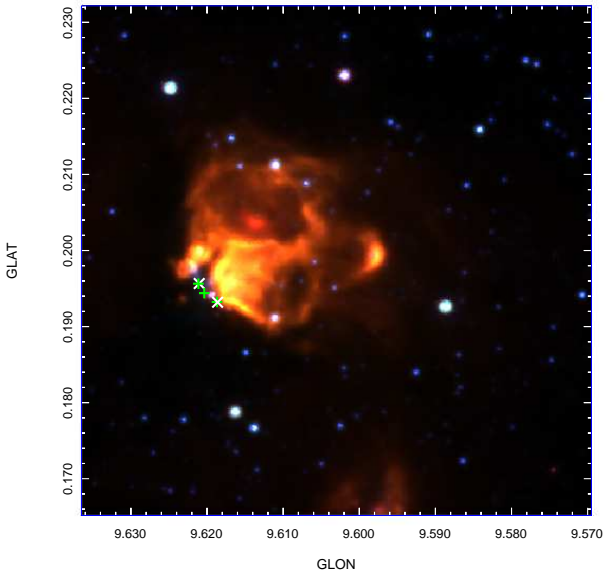}
  %\end{center}
  \caption{\sffamily
The three-color (4.5 $\mu{}m$: blue; 5.8 $\mu{}m$: green; 8.0 $\mu{}m$: red) image of the bubble CN135 from CH07 catalogue, with three OH masers (plus) located on the border of the bubble and two methanol masers (cross) sharing the same sites with their associated OH masers.
   \label{fig:fig10}}
\end{figure}
%\efi
%%%%%%%%%%%%%%%%%%%%%%%%%%%%%%%

% fig11
%%%%%%%%%%%%%%%%%%%%%%%%%%%%%%%
\begin{figure}
  \begin{center}
    \includegraphics*[height=0.5\textwidth,viewport=300 20 900 650]{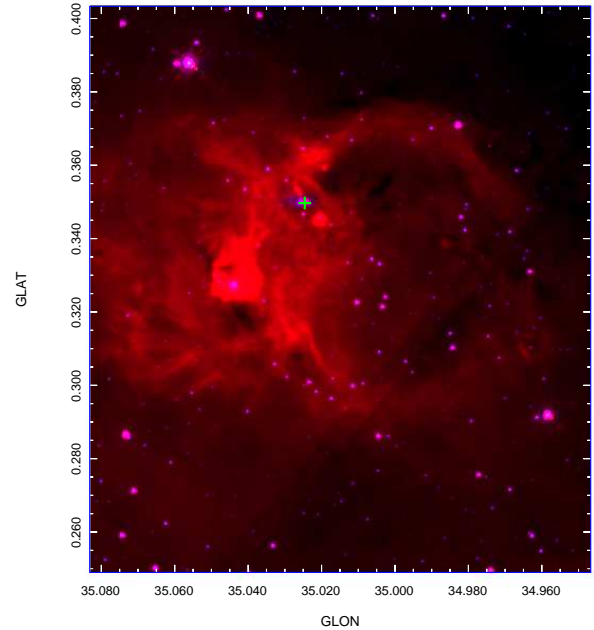}
  \end{center}
  \caption{\sffamily
Same as Fig. \ref{fig:fig10}, but for the bubble N65 from CH06 catalogue. Notably only one OH maser (plus) is located on the border of the bubble.
   \label{fig:fig11}}
\end{figure}
%\efi
%%%%%%%%%%%%%%%%%%%%%%%%%%%%%%%

The \emph{Spitzer}-GLIMPSE images revealed a large number of full or partial ring-like structures, which were referred to as bubbles (\cite{Che2006}; \cite{Che2007}). Bubbles are a common phenomenon in the ISM. Most of them may be produced by newly formed massive stars and clusters, which excite PAHs in the swept up shell, and the PAHs are strong emitters at 8 $\mu{}m$ in the photo-dissociation regions (PDRs) surrounding the H{\sc ii}  region (\cite{Dee2010}). Churchwell et al. (2006, 2007) (hereafter CH06, CH07) have catalogued about 600 ring structures traced mainly by 8.0 $\mu{}m$ emission between Galactic longitudes $-65^{\circ}$ to $65^{\circ}$ by inspecting the GLIMPSE I/II mosaic visually. Deharveng et al. (2010) studied a gallery of bubbles mainly from Churchwell et al. (2006), finding that 86\% of their bubbles enclose H{\sc ii}  regions. Good correlation of MIR bubbles with known H{\sc ii}  regions or radio-continuum emission at 20 cm, and relatively low contamination from asymptotic giant brunch (AGB) star bubbles, supernova remnants (SNRs) and planetary nebulae (PNe) reported in the literature, indicates that bubbles are a good tracer of star formation activity (\cite{Che2006}; \cite{Dee2010}). Deharveng et al. (2010) also concluded that more than a quarter of the bubbles may have triggered the formation of massive objects. The majority studies into triggered star formation by the expansion of bubbles take advantage of multiwavelength data sets (typically with near-, mid-, far-infrared, millimeter and radio wavelengths) to estimate the mass, age, and luminosity of triggering and triggered sources, and compare the kinematic properties of the young stars with the surrounding ISM. The evidence of triggered star formation has thus been reported in several known H{\sc ii}  regions and bubbles, e.g. Sh2-212 (\cite{Dee2008}), W51a (\cite{Kae2009}), RCW120 (\cite{Zae2010}), S51 (\cite{Zhe2012}), N4 (\cite{Lie2013}), and G52L (\cite{Gue2013}). Although these studies provide reasonable evidence of triggered star formation, they always conclude with uncertainties and open questions such as the uncertainties of the association between YSOs and the collected material (\cite{Kae2009}) and the age uncertainties of stars in the condensation regions (\cite{Zae2010}). \cite{Kee2012} took a statistical approach to investigate the association of bubbles with massive star formation and found a strong positional correlation of massive young stellar objects (MYSOs) and bubbles. However, it is yet not clear whether the expansion of bubbles could cause the following generation of stars. Interstellar OH masers trace the massive star formation, thus, the association study between interstellar OH masers and bubbles may obtain the indirect evidence supporting the triggered star formation by the expansion of bubbles.

We use the catalogue made by Churchwell et al. (2006, 2007) which contains 591 bubbles to study the association between bubbles and 219 interstellar OH masers mentioned above. Churchwell et al. (2006, 2007) measured the bubbles' parameters such as their semimajor ($R_{out}$) and semiminor ($r_{out}$) axes of the outer ellipse. We use $1.2R_{out}$ as the criterion to cross-match the OH maser positions and the bubble center positions. The criteria is larger than $R_{out}$ because the definition of $R_{out}$ is subjective. We find that 18 bubbles are associated with 22 OH masers. Among them, one bubble (CN135) is associated with three OH masers, and two bubbles (CN71, S66) are associated with two OH masers, respectively. These associations may be caused by merely geometric effects, but still need to be investigated. Here we assume all the masers are associated with bubbles. Then we cross-match these 22 OH masers with 707 MMB masers taking 2\arcsec\ criterion. The result is that ten methanol masers are associated with ten OH masers and nine bubbles. The basic information about 18 maser-associated bubbles and the names of the associated OH masers and methanol masers are also listed in Table 6.

The low association between bubbles and interstellar OH masers may be due to the maser sample we used. The 219 OH masers as described above are mainly from targeted surveys. Therefore, many bubbles may not have been searched for OH masers on the boarders. In addition, the low association may suggest that the young stars on the boarders of the majority of the bubbles are at an early stage and have not developed the physical conditions suitable for the pumping of OH masers. Besides, the result may simply imply that the majority of the bubbles have not yet triggered the star formation on the borders, or the triggered star formation is inefficient on the borders of bubbles.

We display the false color images of the bubbles (4.5 $\mu{}m$: blue; 5.8 $\mu{}m$: green; 8.0 $\mu{}m$: red) using the display program ds9\footnote{See http://hea-www.harvard.edu/RD/ds9/ref} and point out the positions of OH masers and methanol masers in the bubble infrared images. As examples, Fig. \ref{fig:fig10} shows the three-color image of the bubble CN135. It shows that three OH masers are located on the border of the bubble and two methanol masers share the same sites with OH masers. Fig. \ref{fig:fig11} is the three-color image of the bubble N65. It only has one OH maser on the boarder of the bubble. The remaining sixteen bubbles have the same maser distribution as CN135 or N65, and many masers are associated with bright emission at 8.0 $\mu{}m$. The 6.7 GHz Class II methanol masers and interstellar OH masers are the tracers of massive star formation. This result confirms that the massive star formation is ongoing on the border of bubbles and the bubbles may trigger the massive star formation by their outward expansion.

% fig12
%%%%%%%%%%%%%%%%%%%%%%%%%%%%%%%
\begin{figure}
  %\begin{center}
    \includegraphics[height=0.3\textwidth,width=0.5\textwidth, angle=0]{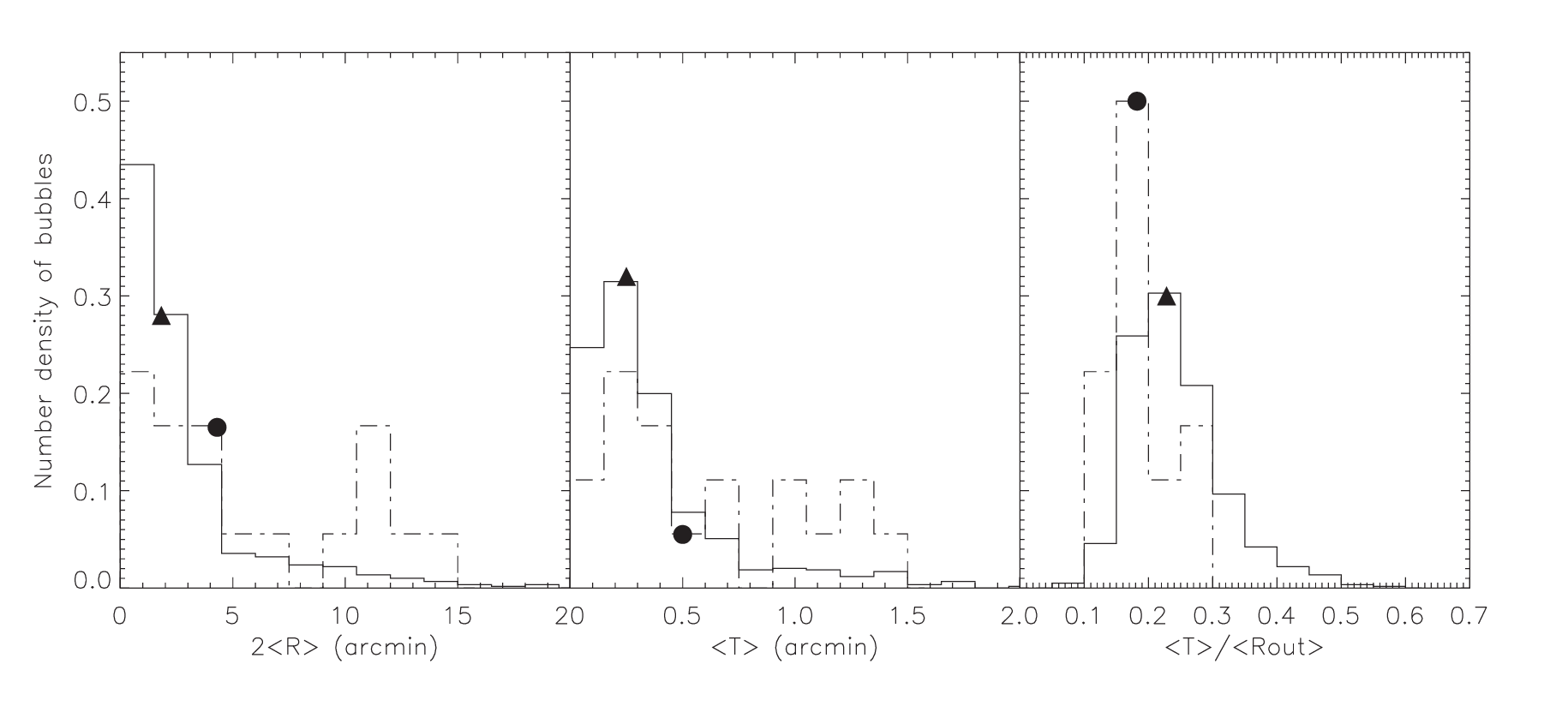}
  %\end{center}
  \caption{\sffamily
Observed distributions of average bubble angular diameters ($2\langle{R}\rangle$), average bubble thickness (angular measure $\langle{T}\rangle$) and bubble thickness relative to average outer radius ($\langle{T}\rangle/\langle{R_{out}}\rangle$). The dot dashed curves show the distributions of 18 maser-associated bubbles, and the solid curves are those of the 591 bubbles. The solid circles are the medians of 18 maser-associated bubbles, while the solid triangles are the medians of the 591 bubbles.
   \label{fig:fig12}}
\end{figure}
%\efi
%%%%%%%%%%%%%%%%%%%%%%%%%%%%%%%

% fig13
%%%%%%%%%%%%%%%%%%%%%%%%%%%%%%%
%\begin{figure}
  %\begin{center}
 %   \includegraphics[height=0.38\textwidth,width=0.48\textwidth, angle=0]{hist-2.ps}
%  %\end{center}
 % \caption{\sffamily
%Distribution of average bubble thickness (angular measure $\langle{T}\rangle$). The dashed curve shows the distribution of 18 maser-associated bubbles, and the solid curve is that of the 591 bubbles.
%   \label{fig:fig13}}
%\end{figure}
%\efi
%%%%%%%%%%%%%%%%%%%%%%%%%%%%%%%

% fig14
%%%%%%%%%%%%%%%%%%%%%%%%%%%%%%%
%\begin{figure}
  %\begin{center}
  %  \includegraphics[height=0.38\textwidth,width=0.48\textwidth, angle=0]{hist-3.ps}
  %\end{center}
 % \caption{\sffamily
%The distribution of bubble thickness relative to average outer radius ($\langle{T}\rangle/\langle{R_{out}}\rangle$). The dashed curve shows the distribution of 18 maser-associated bubbles, and the solid curve is that of the 591 bubbles.
%   \label{fig:fig14}}
%\end{figure}
%\efi
%%%%%%%%%%%%%%%%%%%%%%%%%%%%%%%

Beyond that, we study the properties of maser-associated bubbles. The left figure in Fig. \ref{fig:fig12} is the average angular diameter ($2\langle{R}\rangle$) distribution of 591 bubbles and 18 maser-associated bubbles. We can see from it that the average angular diameters of 18 maser-associated bubbles are slightly larger than that of 591 bubbles. The middle figure in Fig. \ref{fig:fig12} shows that there is a deviation of the average angular thickness ($\langle{T}\rangle$) between 18 maser-associated bubbles and 591 bubbles. From the right figure in Fig. \ref{fig:fig12}, we can see that the ratio of average thickness to average outer radius ($\langle{T}\rangle/\langle{R_{out}}\rangle$) is smaller than 0.3 for 18 maser-associated bubbles. The $2\langle{R}\rangle$, $\langle{T}\rangle$, and $\langle{T}\rangle/\langle{R_{out}}\rangle$ medians for 591 bubbles and 18 maser-associated bubbles are labeled in Fig. \ref{fig:fig12}. The $2\langle{R}\rangle$, and $\langle{T}\rangle$ medians of 18 maser-associated bubbles are 4.30\arcmin\ and 0.50\arcmin, which are larger than 1.82\arcmin\ and 0.25\arcmin\ for 591 bubbles; The $\langle{T}\rangle/\langle{R_{out}}\rangle$ medians for 18 maser-associated bubbles and 591 bubbles are 0.182 and 0.228, respectively. Deharveng et al. (2010) suggest that the large size of bubbles correspond to an older age or evolution in the ISM with low densities. Churchwell et al. (2006) concluded that bubble shell thickness increases approximately linearly with shell radius. The measurements of larger average bubble angular diameter ($2\langle{R}\rangle$) and smaller thickness relative to average outer radius ($\langle{T}\rangle/\langle{R_{out}}\rangle$) of 18 maser-associated bubbles may suggest that the maser-associated bubbles are generally older than normal bubbles. However, these results may be simply due to the artifacts of small number statistics and need further studies.

\section{Summary}

In this paper, we present the catalogues of all the detected interstellar OH masers of different transitions. We also used the GLIMPSE and \emph{WISE} data to investigate the MIR environment of the interstellar OH masers, and explore the MIR environment differences between OH masers and methanol masers. Besides, we also studied the association between OH masers and bubbles.

For maser action, one must have the presence of OH (from the dissociation of water vapor), intense IR radiation (to populate higher energy levels of OH which then cascade to lower states which are inverted by a combination of IR and collisions), and a long path where the velocity dispersion is small, so the maser intensity can rise to a level at which can be detected. An assumption is that the masers radiate have spherical shapes, so radiate in all directions. That is, the masers are not beamed in a certain direction.We found no obvious difference in the GLIMPSE color-color and color-magnitude diagrams for solitary OH masers, OH masers associated with methanol masers, and solitary methanol masers, and this result is different from previous studies (Ellingsen 2006). However, in the \emph{WISE} color-magnitude diagram, the 22 $\mu{}m$ magnitude of MIR counterparts of solitary OH masers is significantly brighter than that of solitary methanol masers. This may indicate that the stellar mass range associated with OH masers can extend to higher masses than that for methanol masers, or it may be because the OH masers trace a later evolutionary phase than methanol masers. For bubbles, 22 OH masers are associated with 18 bubbles. Among them, ten OH masers are also associated with Class II methanol masers. The OH masers and methanol masers are usually found on the border of the bubbles, suggesting there are ongoing star formation activities. This result is the indirect evidence of the triggered star formation by the expansion of bubbles.

%\item[Acknowledgements]
\section*{Acknowledgments}

We thank the refree for the helpful comments and constructive suggestions. This work is based in part on observations made with the Spitzer Space Telescope, which is operated by the Jet Propulsion Laboratory, California Institute of Technology under a contract with National Aeronautics and Space Administration (NASA). This publication makes use of data products from the Wide-field Infrared Survey Explorer, which is a joint project of the University of California Institute of Technology, funded by NASA. We would like to thank the Key Laboratory of Radio Astronomy, Chinese Academy of Sciences. This work is partly supported by China Ministry of Science and Technology under State Key Development Program for Basic Research (2012CB821800), the National Natural Science Foundation of China (grants 10625314, 11073041, 11103006, 11121062, 11133008,  and 11273043), the Knowledge Innovation Program of the Chinese Academy of Sciences (Grant No. KJCX1-YW-18), the Scientific Program of Shanghai Municipality (08DZ1160100), the CAS/SAFEA International Partnership Program for Creative Research Teams, the Strategic Priority Research Program ``The Emergence of Cosmological Structures" of the Chinese Academy of Sciences, Grant No. XDB09000000, and the Shanghai Tian Ma Telescope Pre-research Funding.

\bsp

%\bibliography{mn-jour,paper2}

\begin{thebibliography}{}

%     [\protect\citeauthoryear{}{}]
%-------------------------------------------------------

\bibitem[\protect\citeauthoryear{Anderson et al. 2012}{}]
{Ane2012}Anderson L. D., Zavagno A., Barlow M. J. et al., 2012, A\&A, 537, 1

\bibitem[\protect\citeauthoryear{Argon et al. (2000)}{}]
{Are2000}Argon A. L., Reid M. J., Karl M. Menten., 2000, ApJS, 129, 159, (A2000)

\bibitem[\protect\citeauthoryear{Bains et al. (2008)}{}]
{Bae2008}Bains I., Caswell J., Richards A. M. S. et al., 2008, IAU, 242, 2007

\bibitem[\protect\citeauthoryear{Baudry (1974)}{}]
{Ba1974}Baudry A., 1974, A\&A, 33, 381

\bibitem[\protect\citeauthoryear{Baudry et al. (1997)}{}]
{Bae1997}Baudry A., Desmurs J. F., Cohen R. J., 1997, A\&A, 325, 255, (BD97)

\bibitem[\protect\citeauthoryear{Baudry \& Desmurs (2002)}{}]
{BD2002}Baudry A., Desmurs J. F., 2002, A\&A, 394, 107, (BD2002)

\bibitem[\protect\citeauthoryear{Benjamin et al. 2003}{}]
{Bee2003}Benjamin R. A. et al., 2003, PASP, 115, 953

\bibitem[\protect\citeauthoryear{Caswell et al. 1980}{}]
{Cae1980}Caswell J. L., Haynes R. F., Goss W. M., 1980, AuJPh, 33, 639

\bibitem[\protect\citeauthoryear{Caswell \& Haynes 1983}{}]
{CH1983}Caswell J. L., Haynes R. F., 1983, AuJPh, 36, 361

\bibitem[\protect\citeauthoryear{Caswell \& Haynes 1987}{}]
{CH1987}Caswell J. L., Haynes R. F., 1987, AuJPh, 40, 215

\bibitem[\protect\citeauthoryear{Caswell \& Vaile (1995)}{}]
{CV1995}Caswell J. L., Vaile R. A., 1995, MNRAS, 273, 328, (CV95)

\bibitem[\protect\citeauthoryear{Caswell (1997)}{}]
{Ca1997}Caswell J. L., 1997, MNRAS, 289, 203

\bibitem[\protect\citeauthoryear{Caswell (1998)}{}]
{Ca1998}Caswell J. L., 1998, MNRAS, 297, 215, (C98)

\bibitem[\protect\citeauthoryear{Caswell (1999)}{}]
{Ca1999}Caswell J. L., 1999, MNRAS, 308, 683, (C99)

\bibitem[\protect\citeauthoryear{Caswell 2001}{}]
{Ca2001}Caswell J. L., 2001, MNRAS, 326, 805, (C2001)

\bibitem[\protect\citeauthoryear{Caswell (2003)}{}]
{Ca2003}Caswell J. L., 2003, MNRAS, 341, 551, (C2003)

%\bibitem[\protect\citeauthoryear{Caswell (2004a)}{}]
%{Ca2004a}Caswell J. L., 2004, MNRAS, 349, 99

\bibitem[\protect\citeauthoryear{Caswell (2004)}{}]
{Ca2004}Caswell J. L., 2004, MNRAS, 352, 101, (C2004)

\bibitem[\protect\citeauthoryear{Caswell et al. (2010)}{}]
{CaK2010}Caswell J. L., Kramer B. H., Sukom A. et al., 2010, MNRAS, 402, 2649

\bibitem[\protect\citeauthoryear{Caswell et al. 2010}{}]
{Cae2010}Caswell J. L., Fuller G. A., Green J. A. et al., 2010, MNRAS, 404, 1029

\bibitem[\protect\citeauthoryear{Caswell et al. 2011}{}]
{Cae2011}Caswell J. L., Fuller G. A., Green J. A. et al., 2011, MNRAS, 417, 1964

\bibitem[\protect\citeauthoryear{Charnley et al. 1992}{}]
{Che1992}Charnley S. B., Tielens A. G. G. M., Millar T. J., 1992, ApJ, 399, L71

\bibitem[\protect\citeauthoryear{Charnley et al. 1995}{}]
{Che1995}Charnley S. B., Kress M. E., Tielens A. G. G. M. et al., 1995, ApJ, 448, 232

\bibitem[\protect\citeauthoryear{Churchwell et al. 2006}{}]
{Che2006}Churchwell E., Povich M. S., Allen D. et al., 2006, ApJ, 649, 759

\bibitem[\protect\citeauthoryear{Churchwell et al. 2007}{}]
{Che2007}Churchwell E., Watson M. S., Povich M. S. et al., 2007, ApJ, 670, 428

\bibitem[\protect\citeauthoryear{Cohen et al. (1991)}{}]
{Coe1991}Cohen R. J., Masheder M. R. W., Walker R. N. F., 1991, MNRAS, 250, 611, (CM91)

\bibitem[\protect\citeauthoryear{Cohen et al. (1995)}{}]
{Coe1995}Cohen R. J., Masheder M. R. W., Caswell J. L., 1995, MNRAS, 274, 808, (CM95)

\bibitem[\protect\citeauthoryear{Cragg et al. (2002)}{}]
{Cre2002}Cragg D. M., Sobolev A. M., Godfrey P. D., 2002, MNRAS, 331, 521

\bibitem[\protect\citeauthoryear{Dartois et al. 1999}{}]
{Dae1999}Dartois E., Schutte W., Geballe T. R. et al., 1999, A\&A, 342, L32

\bibitem[\protect\citeauthoryear{Deharveng et al. 2008}{}]
{Dee2008}Deharveng L., Lefloch B., Kurtz S. et al., 2008, A\&A, 482, 585

\bibitem[\protect\citeauthoryear{Deharveng et al. 2010}{}]
{Dee2010}Deharveng L., Schuller F., Anderson L. D. et al., 2010, A\&A, 523, 6

\bibitem[\protect\citeauthoryear{Dodson \& Ellingsen (2002)}{}]
{DE2002}Dodson R. G., Ellingsen S. P., 2002, MNRAS, 333, 307, (DE2002)

\bibitem[\protect\citeauthoryear{Draine 2003}{}]
{Dr2003}Draine B. T., 2003, ARA\&A, 41, 241

\bibitem[\protect\citeauthoryear{Edris et al. (2007)}{}]
{Ede2007}Edris K. A., Fuller G. A., Cohen R. J., 2007, A\&A, 465, 865, (EF2007)

\bibitem[\protect\citeauthoryear{Ellingsen et al. 1996}{}]
{Ele1996}Ellingsen S. P., von Bibra M. L., McCulloch P. M. et al., 1996, MNRAS, 280, 378

\bibitem[\protect\citeauthoryear{Ellingsen (2006)}{}]
{El2006}Ellingsen S. P., 2006, ApJ, 638, 241

\bibitem[\protect\citeauthoryear{Fish \& Sjouwerman (2007)}{}]
{Fie2007}Fish V. L., Sjouwerman L. O., 2007, ApJ, 668, 331


\bibitem[\protect\citeauthoryear{Forster \& Caswell (1999)}{}]
{FC1999}Forster J. R., Caswell J. L., 1999, A\&A, 137, 43, (FC99)

\bibitem[\protect\citeauthoryear{Forster \& Caswell (1989)}{}]
{FC1989}Forster J. R., Caswell J. L., 1989, A\&A, 213, 339

\bibitem[\protect\citeauthoryear{Gardner \& Martin (1983)}{}]
{GM1983}Gardner F. F., Mart\'{\i}n-Pintado J., 1983, A\&A, 121, 265

\bibitem[\protect\citeauthoryear{Green et al. 1997}{}]
{Gre1997}Green A. J., Frail D. A., Goss W. M. et al., 1997, AJ, 114, 2058

\bibitem[\protect\citeauthoryear{Green et al. 2010}{}]
{Gre2010}Green J. A., Caswell J. L., Fuller G. A. et al., 2010, MNRAS, 409, 913

\bibitem[\protect\citeauthoryear{Green et al. 2012}{}]
{Gre2012}Green J. A., Caswell J. L., Fuller G. A. et al., 2012, MNRAS, 420, 3108

\bibitem[\protect\citeauthoryear{Guilloteau (1982)}{}]
{Gu1982}Guilloteau S., 1982, A\&A, 116, 101

\bibitem[\protect\citeauthoryear{Gundermann (1965)}{}]
{Gu1965}Gundermann E. J., 1965, Ph.D. thesis, Harvard University

\bibitem[\protect\citeauthoryear{Harvey-Smith \& Cohen (2005)}{}]
{HC2005}Harvey-Smith L., Cohen R. J., 2005, MNRAS, 356, 673, (SC2005)

\bibitem[\protect\citeauthoryear{Harvey-Smith \& Cohen (2007)}{}]
{HC2007}Harvey-Smith L., Cohen R. J., 2007, IAU, 237, 2006

\bibitem[\protect\citeauthoryear{Hekkert et al. 1996}{}]
{HC1996}Hekkert P. L., Chapman J. M., 1996, A\&AS, 119, 459

\bibitem[\protect\citeauthoryear{Kang et al. 2009}{}]
{Kae2009}Kang M., Bieging J. H., Kulesa C. A. et al., 2009, ApJ, 701, 454

\bibitem[\protect\citeauthoryear{Kendrew et al. (2012)}{}]
{Kee2012}Kendrew S., Simpson R., Bressert E. et al., 2012, ApJ, 755, 71

\bibitem[\protect\citeauthoryear{Li et al. 2013}{}]
{Lie2013}Li J. Y., Jiang Z. B., Liu Y. et al., 2013, RAA, 13, 921

\bibitem[\protect\citeauthoryear{Li et al. 2013}{}]
{Gue2013}Li G. X., Wyrowski F., Menten K. et al., 2013, arXiv:1310.3267

%\bibitem[Lewis et al (1995)]{Lewis1995}
%Lewis, B. M. \& Engels, D. 1995, MNRAS, 274, 429
\bibitem[\protect\citeauthoryear{Migenes et al. (2005)}{}]
{Mie2005}Migenes V., Cruz-Vazquez L., Slysh V. I. et al., 2005, ASPC, 340, 361

\bibitem[\protect\citeauthoryear{Mu et al. (2010)}{}]
{Mue2010}Mu J. M., Esimbek J., Zhou J. J. et al., 2010, RAA, 10, 166

\bibitem[\protect\citeauthoryear{Palmer et al. (2004)}{}]
{Pae2004}Palmer P., Goss W. M., Whiteoak J. B., 2004, MNRAS, 347, 1164, (PG2004)

\bibitem[\protect\citeauthoryear{Pavlakis \& Kylafis (1996)}{}]
{PK1996}Pavlakis K. G., Kylafis N. D., 1996, ApJ, 467, 309

\bibitem[\protect\citeauthoryear{Pavlakis \& Kylafis 2000}{}]
{PK2000}Pavlakis K. G., Kylafis N. D., 2000, ApJ, 534, 770

\bibitem[\protect\citeauthoryear{Reid 2002}{}]
{Re2002}Reid M. J., 2002, IAUS, 206, 506

\bibitem[\protect\citeauthoryear{Slysh et al. (2010)}{}]
{Sle2010}Slysh V. I., Alakoz A. V., \& Migenes V., 2010, MNRAS, 404, 1121

\bibitem[\protect\citeauthoryear{Smits (1994)}{}]
{Sm1994}Smits D. P., 1994, MNRAS, 269, L11, (S94)

\bibitem[\protect\citeauthoryear{Smits (1997)}{}]
{Sm1997}Smits D. P., 1997, MNRAS, 287, 253, (S97)

\bibitem[\protect\citeauthoryear{Smits (2003)}{}]
{Sm2003}Smits D. P., 2003, MNRAS, 339, 1, (S2003)

\bibitem[\protect\citeauthoryear{Szymczak et al. (2000)}{}]
{Sze2000}Szymczak M., Kus A. J., Hrynek G., 2000, MNRAS, 312, 211, (SK2000)

\bibitem[\protect\citeauthoryear{Szymczak \& G\'{e}rard (2004)}{}]
{SG2004}Szymczak M., G\'{e}rard E., 2004, A\&A, 414, 235, (SG2004)

\bibitem[\protect\citeauthoryear{Thacker \& Wilson (1970)}{}]
{TW1970}Thacker D. L., Wilson W. J., Barrett A. H., 1970, ApJ, 161, L191

\bibitem[\protect\citeauthoryear{Turner et al. (1970)}{}]
{Tue1970}Turner B. E., Palmer P., Zuckerman B., 1970, ApJ, 160, L125


\bibitem[\protect\citeauthoryear{Weaver et al. (1965)}{}]
{Wee1965}Weaver H., Williams R. W., Dieter N. H. et al., 1965, Natur, 208, 29

\bibitem[\protect\citeauthoryear{Whitney et al. (2003)}{}]
{Whe2003}Whitney B. A., Wood K., Bjorkman J. E. et al., 2003, ApJ, 598, 1079

\bibitem[\protect\citeauthoryear{Wright et al. 2010}{}]
{Wre2010}Wright E. L., Eisenhardt P. R. M., Mainzer A. K. et al., 2010, AJ, 140, 1868

\bibitem[\protect\citeauthoryear{Yen et al. (1969)}{}]
{Yee1969}Yen J. L., Zuckerman B., Palmer P. et al., 1969, ApJ, 156, L27

\bibitem[\protect\citeauthoryear{Zavagno et al. 2010}{}]
{Zae2010}Zavagno A., Russeil D., Motte F. et al., 2010, A\&A, 518, L81

\bibitem[\protect\citeauthoryear{Zhang et al. 2012}{}]
{Zhe2012}Zhang C. P., Wang J. J., 2012, A\&A, 544, A11

\bibitem[\protect\citeauthoryear{Zuckerman et al. 1968}{}]
{Zue1968}Zuckerman B., Palmer P., Penfield H. et al., 1968, ApJ, 153, L69

%----------
\end{thebibliography}

%\appendix
%\section[]{Large gaps in L\lowercase{y}${\balpha}$ forests\\* due to fluctuations in line distribution}

\clearpage
\onecolumn

\section*{Appendix : Online material}
\scriptsize
\begin{longtable}{lllllcccll}

%      \begin{minipage}{105mm}
     \caption*{\textbf{Table 2.}
Catalogue of 18 cm Interstellar OH masers.}

\\
\hline
\hline

No. &Source name & RA(J2000) & DEC(J2000)  & Freq.     & Vpeak         & Vrange      & Flux density&  Ref.  & Other name \\
 &            &   (h  m  s)        &($^{o}$  '  ")             & (MHz)     & (km s$^{-1}$) & (km s$^{-1}$) &   (Jy)&   &          \\
\hline

1 & G126.715-0.822 & 01 23 33.17 & +61 48 49.2 & 1665L & -12.06 & -12.06,-8.64 & $>$1.50 & A2000 & \\
1 & G126.715-0.822 & 01 23 33.17 & +61 48 49.2 & 1665R & -5.74 & & $>$0.50 & A2000 & \\
2 & G133.715+1.215 & 02 25 40.59 & +62 05 50.5 & 1665L & -39.29 & -43.96,-37.59 & $>$4.31 & A2000 & W3 \\
2 & G133.715+1.215 & 02 25 40.59 & +62 05 50.5 & 1665R & -39.2 & & $>$0.75 & A2000 & W3 \\
3 & G133.946+1.064 & 02 27 03.70 & +61 52 25.4 & 1612L & -43.15 & -47.51,-42.3 & $>$1.68 & A2000 & W3 OH \\
\hline
\end{longtable}

\scriptsize
\begin{longtable}{lllllcccll}

%\begin{minipage}{105mm}
\caption*{\textbf{Table 3.}
Catalogue of 6 cm Interstellar OH masers.}

\\
\hline
\hline

No. &Source name & RA(J2000) & DEC(J2000)  & Freq.     & Vpeak         & Vrange      & Flux density&  Ref.  & Other name \\
       &  &   (h  m  s)        &($^{o}$  '  ")             & (MHz)     & (km s$^{-1}$) & (km s$^{-1}$) &   (Jy)&   &          \\
\hline

1 & W3-IRS5 & 02 25 40.7716 & +62 05 54.441 & 4765 & -38.95 &  & 0.609 & SC2005 & IRAS02219+6152 \\
2 & W3(OH) & 02 27 03.7129 & +61 52 24.641 & 4765 & -45.01 &  & 72.218 & SC2005 & IRAS02232+6138 \\
3 & ORION-IR & 05 35 14.48 & -05 22 26.6 & 4660 & 14.4 &  &  & GM83 &  \\
4 & MonR2 & 06 07 47.8403 & -06 22 56.576 & 4765 & 10.62 &  & 7.722 & SC2005 & S252-A3 \\
5 & IRAS06055+2039 & 06 08 32.8 & +20 39 16 & 4750 & 10 &  & 1 & CM91 & S252-A3 \\
\hline
\end{longtable}

\scriptsize
\begin{longtable}{lllllcccll}

%\begin{minipage}{105mm}
\caption*{\textbf{Table 4.}
Catalogue of 5 cm Interstellar OH masers.}

\\
\hline
\hline

No. &Source name & RA(J2000) & DEC(J2000)  & Freq.     & Vpeak         & Vrange      & Flux density&  Ref.  & Other name \\
       &  &   (h  m  s)        &($^{o}$  '  ")             & (MHz)     & (km s$^{-1}$) & (km s$^{-1}$) &   (Jy)&   &          \\
\hline

1 & IRAS02232+6138 & 02 27 03.88 & +61 52 24.6 & 6031 & -43.33 & -47.30,-41.89 & 49.3 & BD97 & W3(OH) \\
1 & IRAS02232+6138 & 02 27 03.88 & +61 52 24.6 & 6035 & -43.25 & -48.71,-41.9 & 115.4 & BD97 & W3(OH) \\
2 & G208.99-19.38 & 05 35 14.4 & -05 22 25 & 6035 &  & 3,23 & 0.29 & CV95 &  \\
3 & IRAS05392-0214 & 05 41 44.8 & -02 13 23 & 6035 & 11.07 & 11.07,11.15 & 0.2 & BD97 &  \\
4 & G240.32+0.07 & 07 44 53.2 & -24 07 39 & 6035 &  & 63,64 & 2.3 & CV95 &  \\
\hline
\end{longtable}

\begin{longtable}{lllllcccll}

%\begin{minipage}{105mm}
\caption*{\textbf{Table 5.}
Catalogue of 2.3 cm Interstellar OH masers.}

\scriptsize
\\
\hline
\hline

No. &Source name & RA(J2000) & DEC(J2000)  & Freq.     & Vpeak         & Vrange      & Flux density&  Ref.  & Other name \\
  &       &   (h  m  s)        &($^{o}$  '  ")             & (MHz)     & (km s$^{-1}$) & (km s$^{-1}$) &   (Jy)&   &          \\
\hline

1 & W3(OH) & 02 27 03.88 & +61 52 24.57 & 13441 & -43.03 &  & 45.3 & BD2002 & IRAS02232+6138 \\
1 & W3(OH) & 02 27 03.88 & +61 52 24.57 & 13435L & -45.22 &  &  & BD2002 &  \\
2 & G309.921+0.479 & 13 50 41.77 & -61 35 10.1 & 13441 & -57.94 &  & 6.33 & C2004 &  \\
3 & G311.596-0.398 & 14 06 18.35 & -62 00 15.3 & 13441 & 30.43 &  & 0.48 & C2004 &  \\
4 & G328.809+0.633 & 15 55 48.73 & -52 43 05.5 & 13441 & -44.74 &  & 0.46 & C2004 &  \\
5 & G329.339+0.148 & 16 00 33.15 & -52 44 39.8 & 13441 & -104.44 &  & 0.31 & C2004 &  \\
6 & G331.542-0.066 & 16 12 09.01 & -51 25 47.7 & 13441 & -85.21 &  & 0.29 & C2004 &  \\
7 & G336.941-0.156 & 16 35 55.20 & -47 38 45.4 & 13441 & -65.22 &  & 0.36 & C2004 &  \\
8 & G345.003-0.224 & 17 05 11.20 & -41 29 07.0 & 13441 & -28.25 &  & 0.18 & C2004 &  \\
9 & G11.904-0.141 & 18 12 11.44 & -18 41 29.0 & 13441 & 43.00  &  & 0.31 & C2004 &  \\
10 & IRAS20081+3122 & 20 10 09.14 & +31 31 34.37 & 13441 & -0.31 &  & 0.95 & BD2002 & ON1 \\
11 & IRAS21413+5442 & 21 43 01.36 & +54 56 16.28 & 13441 & -61.47 &  & 0.77 & BD2002 &  \\

\hline

\end{longtable}

\label{lastpage}

\end{document}